\documentclass[%
 reprint,
 superscriptaddress,
 showpacs,
  nofootinbib,
 amsmath,amssymb,
 aps,
prc,
]{revtex4-1}

\usepackage{verbatim}
\usepackage{graphicx}
\usepackage{epsfig}
\usepackage{amsfonts}
\usepackage{amsmath,amssymb}
\usepackage{bm}
\usepackage[colorlinks=true,citecolor=blue,urlcolor=blue,linkcolor=blue,menucolor=blue]{hyperref}
\hypersetup{breaklinks=true}
\usepackage{color}

\begin{document}

\title{Calculation of expectation values of operators in the Complex Scaling method}

\author{G. Papadimitriou}
\email{papadimitrio1@llnl.gov}
\affiliation{
Nuclear and Chemical Science Division, Lawrence Livermore National Laboratory, Livermore, CA 94551, USA
}%



\begin{abstract}

The complex scaling method (CSM) provides with a way to obtain resonance parameters of particle unstable states by rotating the coordinates and momenta
of the original Hamiltonian. It is  convenient to use an L$^2$ integrable basis to resolve the complex rotated or complex scaled Hamiltonian H$_{\theta}$,
with $\theta$ being the angle of rotation in the complex energy plane.
Within the CSM, resonance and scattering solutions do not exhibit an outgoing or scattering wave asymptotic behavior, but rather have decaying asymptotics. 
One of the consequences is that,  expectation values of operators in a resonance or scattering  complex scaled solution
 are calculated by complex rotating the operators. 
 In this work we are exploring applications of the CSM on calculations of expectation values of quantum mechanical operators by
 retrieving the Gamow asymptotic character of the decaying state and calculating hence the expectation value using the unrotated
 operator. The test cases involve a schematic two-body  Gaussian model and also applications
 using realistic interactions.

\end{abstract}

\pacs{21.45.-v,21.45.Bc,21.60.De,24.10.-i}

\maketitle



\section{Introduction} 
\label{intro}

When a nucleus or any other quantum mechanical system is in a metastable state with positive energy
 and decays to a more stable configuration by emitting massive particles or photons, the state is 
  widely known as a resonance. For the resonance to be fully characterized one needs to know the positive
 energy above the associated threshold, which can be denoted as E$_r$, and also a quantity that would be related to the time that is takes for the 
 resonance to decay to the more stable configuration. The latter is known
 as the width $\Gamma$ of the resonance. E$_r$ and $\Gamma$ uniquely define the resonance 
 and they are called resonance parameters.  
 
 Resonances appear in reaction experiments,  manifested as enhanced ``bumps" in the measured cross-sections.
  In nuclear physics,  resonance parameters have been extracted  by fitting
 the Thomas-Lane formulas \cite{thomas_lane} to the data, as for example in the compilation of light nuclei in \cite{tilley1992energy}. This process is known as phenomenological R-matrix and has
 been the workhorse for data evaluation and determination of resonance parameters.   
 There also exists a wealth of microscopic methods to solve the nuclear many-body problem and obtain resonance parameters.
 At this point one needs to distinguish the phenomenological R-matrix for fitting the data, to the calculable R-matrix \cite{Descouvemont,fresco}.

 The calculable R-matrix provides with a numerical foundation for solving the scattering problem by assuming a partition of the space into
 internal (bound states regime) and external (scattering regime) and using a matching of the solutions and appropriate boundary conditions at infinity.
 It has been applied for evaluation of scattering observables and resonances with both phenomenological  \cite{Descouvemont,fresco} and also realistic nucleon-nucleon interactions \cite{quagli}.
 There has been a lot of work for describing nuclear physics phenomena with positive energies  and the effort cannot be captured in this work, howewer
 it needs to be mentioned that another basic numerical tool to study reaction mechanisms is the so-called continuum discretized coupled channel (CDCC)
 approach, especially for multi-channel reactions and reactions that involve reactions of nuclei with fragile radioactive beams \cite{cdcc,thompson}.   

The last decade there is a revival of methods that are developed in the complex energy plane and provide
a solid theoretical framework towards the unification of structure and reaction aspects of nuclei.
Examples are calculations in the Berggren basis \cite{review_GSM,betan,ncgsm,gaute_michel,forssen,yannen} which is a complex single particle (s.p.) basis that utilizes  extended completeness
relations \cite{Berggren1968265} and unifies resonant and non resonant continuum degrees of freedom. \footnote{When working in the complex energy plane we will use the word resonant that defines
both bound states and resonances and non-resonant continuum which is basically a complex scattering state.}

A resonant state or Gamow state or complex pole of the S-matrix, is a state regular at the origin that satisfies
outgoing boundary conditions at infinity. In this sense, resonant states cannot be described by Hermitian Quantum Mechanics (QM)
which deal with wavefunctions that vanish at large distances. However, since very soon the usefulness of the resonant states was realized for the theoretical description of time dependent processes (e.g. radioactive decays)
and Open Quantum Systems (OQSs) \cite{Okolowicz2003271},
solid  mathematical foundations were developed  \cite{regge,abc1,*abc2,simon_abc,delamadrid,civitarese} in the framework of non-Hermitian QM and also advances in numerical techniques and non-Hermitian diagonalizers 
were called for \cite{complex_lanczos_csm,non_hermitian_clusters,Mizusaki01092014,GPLHR}.  

Studying reactions in the complex energy plane can have some attractive advantages. It has been observed that even in the phenomenological R-matrix case, it is beneficial to
perform a continuation of the S-matrix in the complex energy plane in order to determine more reliably resonance parameters \cite{PhysRevC.55.536}. 
At first glance the complex energy would seem unnatural since reactions take place on the
real energy axis. The formulation on the complex energy, however, offers a mathematical getaway that alleviates some, difficult to tackle, problems on the real axis
and then an analytical continuation on the real axis takes place in order to calculate scattering quantities. 

On the real axis and on the configuration space it is known that boundary conditions  quickly become complicated
by increasing the number of reactive particles and in the case of three charged particles the asymptotic is not even known in closed form. 
Such boundary conditions are not appearing in momentum space, but then one deals with singularities which are treated
by calculating the resolvent in the complex energy plane by going over the singularity by small finite radii \cite{kamada,kamada2}. This technique is widely known as complex energy method and was recently employed successfully for 
describing reactions above four-body breakup threshold with realistic interactions \cite{deltuva,*deltuva2} and also for calculations on the lattice \cite{rupak_lee}. The analytical continuation on the real axis
is taken after extrapolations of the radius, that goes over the singularity, to zero. 

The other complex energy alternative is based on a rotation of coordinates and momenta
of the Hamiltonian which leads in bound state like boundary conditions for the description of scattering. The work was originated by Nuttall and Cohen \cite{Nuttall_Cohen}, 
as a way of solving the purely scattering many-body problem for energies above the break-up threshold  to obtain scattering amplitudes without imposing many-body scattering boundary conditions.
It was shown to be successful for both short range and long range potentials \cite{mccurdy_1997,kruppa_scat_ampl,yuma}, for mean field calculations \cite{shi_min}, for scattering calculations above the four body break-up threshold \cite{lazauskas2} and recently  also facilitated modern nuclear forces \cite{Lazau,george1,george2}. 
This complex energy formalism appears with different names and flavors in bibliography such as complex-coordinate method \cite{ccm,Nuttall_Cohen} or complex scaled Lippmann-Schwinger (CSLS) method \cite{kruppa_scat_ampl,yuma}.
In our work we are using the uniform complex rotation of coordinates and momenta which will give rise to a complex eigenvalue problem and in all the following this will be denoted as Complex Scaling Method (CSM) \cite{moiseyev,ykho} which is also the most widely known name.
In this case the analytical continuation on the real axis and the connection with real energy scattering observables is done through the calculation of the complex scaled Green's operator that is used for the evaluation of the continuum level density.

Within the CSM, one is able to describe resonances naturally by calculating  the complex eigenstates
of a complex symmetric non-Hermitian Hamiltonian matrix. The wavefunctions $\Phi_{\theta}$, which are a linear combination of $L^2$ integrable
basis states, do not have the asymptotic behavior that would characterize a resonant state and they fall-off at large distances.
It is known that in order to obtain the Gamow or outgoing character of the CSM wavefunction one needs to perform the backrotation operation: $\Phi$ = U$(\theta)$$^{-1}$$\Phi_{\theta}$. 
The backrotation, however, is shown to be very unstable and it belongs in the category of ill-posed inverse problems.

There are ill-posed inverse problems within the low-energy nuclear physics field that have been tackled through regularization process. The inversion of the Laplace Transform faced
in calculations of the Trento group for electromagnetic response functions  was treated with the Lorentz Inverse Transform ) method \cite{Efros1994130}, while for calculations of nuclear responses within the Argonne's imaginary time Quantum Monte 
Carlo (QMC) methods, the inversion of the imaginary time Euclidean response is  stabilized via maximum entropy techniques (MET) \cite{lovato}.
In QMC calculations of the Seattle-Warsaw groups, the spectral weight function which is calculated by inversion of the Green's function, 
is  also regularized by using the MET \cite {qmc_2009,Magierski20122264} or by singular value decomposition (SVD) \cite{Magierski20122264}.
Coming back to the CSM and the backrotation inverse problem, the regularization solution was proposed in \cite{kruppa_george}, it is known as Tikhonov regularization \cite{tikhonov,Tikh_orig,Tikh_book} and results in retrieving a meaningful Gamow state from the complex scaled solutions.

In this work we will apply the Tikhonov regularization technique for the backrotation of the complex scaled wavefunction and we will calculate expectation values of the 
radius and dipole operators. We will study the behavior of the backrotated wavefunctions and check the stability of the expectation values on the regularization parameter.
We will apply our techniques to a schematic 3D Gaussian problem \cite{gyar_br,Myo20141,Baye20151} which supports bound states and resonances.
We will also calculate the scattering phase shifts of the Gaussian  model and deduce from them resonance parameters using R-matrix formulas. Having
simultaneously the resonance parameters from the diagonalization of the complex scaled Hamiltonian we will study the limits of applicability of traditional R-matrix formulas, especially
when the resonance has a large width.
Finally we will present applications of the Tikhonov technique for the case of the deuteron system and study the expectation value of the dipole operator for the transition from
the $^3$S$_1$-$^3$D$_1$ coupled channel bound ground state (g.s.) to the $^3$P$_1$ continuum states and conclude with perspectives and future plans.

\section{The CSM}
We will briefly mention the basic aspects of the CSM and also refer the reader to \cite{moiseyev,ykho}.
The CSM is mathematically based on the Aguilar-Balslev-Combes (ABC) theorem \cite{abc1,*abc2}. 
A resonance can be revealed in the spectrum of the original Hamiltonian once the
momenta and coordinates of the latter are uniformly rotated in the complex energy plane. The ABC 
theorem then states that the resonant states of the original Hamiltonian are invariant and the non-resonant scattering states are
rotated and distributed on a 2$\theta$ ray that cuts the complex energy plane with a corresponding threshold
being the rotation point. 
 Within the CSM, a resonant state behaves asymptotically as a bound state (see also Fig.\ref{Fig1}), which 
implies that the description of resonances does not require special boundary conditions at infinity and their description adopting
an L$^2$ integrable basis is sufficient.

The Hamiltonian is transformed under the action of the non unitary complex scaling  operator U$(\theta)$ as:
\begin{equation}
\label{cs_hami}
H(\theta,r) = U(\theta)H(r)U(\theta)^{-1},
\end{equation}
where we have assumed for simplicity that the Hamiltonian depends only on the coordinate r and   U$(\theta)$ has the following property when acting on a state:
\begin{equation}
\label{ucs}
U(\theta)\Phi(r) = e^{\frac{3}{2}if\theta}\Phi(re^{i\theta})
\end{equation}
with $f$ = 1 for a two-body system. CSM has the flexibility to be adopted to the coordinate system one uses to describe a 
particular nucleus, hence the degrees of freedom (or f in \eqref{ucs}) is usually larger than one e.g. for multi-cluster systems.
One of the ways to solve the problem governed by the Hamiltonian in Eq.\eqref{cs_hami} is the expansion in a complete basis.
Any L$^2$ basis could be employed for this purpose, such as the Harmonic Oscillator (HO) basis, Slater basis \cite{kruppa_george},tempered Gaussian basis \cite{tempered},
basis defined on a grid such as Lagrange mesh \cite{Baye20151}, Discrete Variable Representation basis \cite{dvr} etc.
We are adopting for the time being the HO basis characterized by a length parameter $b$, without being restrictive to other more flexible basis and
we end up with a complex symmetric eigenvalue problem:
\begin{equation}
\label{zgeev}
H_{\theta}\Phi_{\theta} = E_{\theta}\Phi_{\theta}.
\end{equation}
 
\section{Computations}
\label{computations}

\subsection{Backrotation, expectation values of operators and phase-shifts with a schematic potential}
\label{schematic_pot}

\subsubsection{Backrotation of CSM wavefunction}
\label{backrot}

We start our calculations by considering a schematic Hamiltonian  $H = -\frac{\hbar^2}{2\mu}\nabla^2 + V$ with the potential consisting of two Gaussian form factors, one attractive and one repulsive:
\begin{equation}
\label{toy}
V(r) = -8.0e^{-0.16r^2} + 4.0e^{-0.04r^2}
\end{equation}
and also working in a system of units where $\hbar$ = 1, $b$=1, $\mu$ = 1.
It is worth noting that this potential was employed for first time in calculations for the exploration of the direct backrotation and how one could minimize errors associated with it in \cite{gyar_br} and later on for
CSM calculations of the complex scaled Green operator \cite{Homma01041997,Myo01051998,odsuren}. It also provides with a testing bed for other methods as well \cite{Baye20151}.
\begin{figure}[h!] 
  \includegraphics[width=\columnwidth]{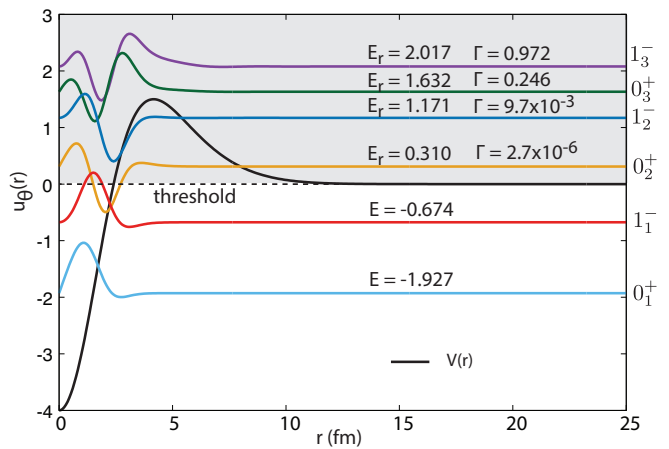} 
  \caption[T]{\label{Fig1}
  (Color online) Resonant eigenvalues and radial wavefunctions of the complex scaled schematic Gaussian Hamiltonian. 
  The gray area denotes the positive energy part of the spectrum above threshold. Even though the last four states are above threshold,
  their radial dependence resembles the one of a bound state. The energies and widths are in MeV.}
\end{figure}
For the Hamiltonian in Eg.\eqref{toy} the transformation \eqref{cs_hami} implies a uniform rotation of the coordinate as $r$ $\to$ $re^{i\theta}$ and momentum $p$ $\to$ $pe^{-i\theta}$.
The kinetic energy becomes $\nabla^2$ $\to$ $\nabla^2e^{-2i\theta}$ and for the local potential we just have to scale the coordinate, so basically $r^2$ $\to$ $r^2e^{2i\theta}$.
The solution is assumed to have the form \eqref{cs_wf} and the complex coefficients are determined by solving the non-Hermitian
complex symmetric eigenvalue problem.
In Fig.\ref{Fig1} we gather some of the resonant solutions that the potential supports for the $\ell$=0,1 states for 
a basis size of N=30 HO radial nodes and a rotation angle $\theta$ = 0.35 radians. The resonant states depicted are
invariant with respect to changes in $\theta$ and convergence was tested as a function of both the basis size and $\theta$. For a range of rotation angles up to $\theta$ = 0.6 radians and for N=30 the real and imaginary part of the resonant states
are unchanged up to the sixth significant digit.

We notice of course that even though the resonant solutions above the threshold are all complex with non-zero widths, their radial dependence is reminiscent
of a bound state. This is to be expected since the radial wavefunctions are expanded in an L$^2$ basis as:
\begin{equation}
\label{cs_wf}
u_{\theta}(r) = \sum_{n=1}^N C_n^{\theta}\phi_{n}(r)
\end{equation}
where the expansion coefficients are the complex eigenvectors of the H$_{\theta}$ diagonalization and $\phi_{n}(r)$ in our
case are the spherical HO radial basis functions. The radial wavefunctions have both real and imaginary parts because of the complex nature of the
C$^{\theta}$ coefficients.

\begin{figure}[t] 
  \includegraphics[width=\columnwidth]{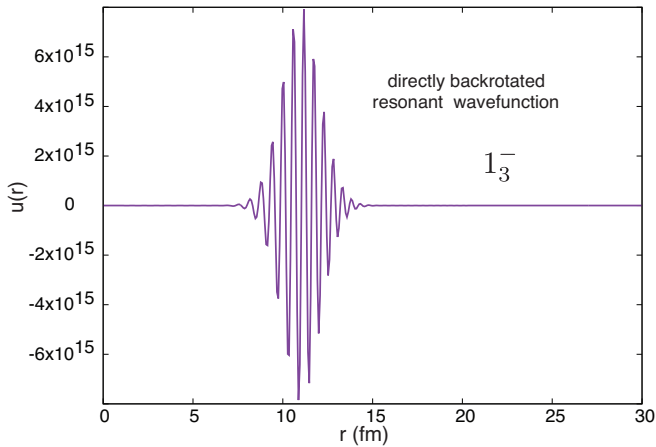} 
  \caption[T]{\label{Fig2}
  (Color online)  Real part of the backrotated 1$_{3}^{-}$ resonant state for $\theta$=0.35 rad and N=30. The subscript $\theta$ was dropped to differentiate from the CSM $u_{\theta}(r)$ solution.}
\end{figure}

In order to obtain the Gamow character of the radial wavefunction u(r), as noted earlier, the backrotation transformation $U(\theta)^{-1}$ has to be applied on the CSM solution.
In the case of the two-body problem and according to \eqref{ucs}   that would constitute to the following transformation:
\begin{equation}
u(r) = e^{-\frac{3}{2}i\theta} \sum_{n=1}^N C_n^{\theta}\phi_n(re^{-i\theta}).
\end{equation}
As it was shown in \cite{kruppa_george} and also presented here in Fig.\ref{Fig2}, the direct backrotation 
is very unstable and retrieving the Gamow character of the wavefunction in this way is not possible.
The source of the problem is two-fold. First, the 
expansion coefficients contain errors which are magnified in the process of the inversion. Second and actually the main
source of the instability is that the HO backrotated basis functions exhibit a very large amplitude oscillatory behavior when increasing
the back rotation angle $\theta$. This is exactly what is depicted in Fig.\ref{Fig2} for the backrotation of the  broad 1$_{3}^{-}$ resonant state.
Eventually the oscillations diminish for large distances, but at intermediate distances we observe very large amplitudes. This behavior of the complex scaled HO basis functions was the reason in
\cite{contour_dm} the authors  worked in momentum space and adopted the contour deformation method for solving a complex  momentum T-matrix equation. We also note that in a recent work of the CSM with realistic
potentials \cite{george1,george2} the backrotation of HO basis functions did not cause any problem since, due to the short range nature of the nucleon-nucleon force, the complex scaled matrix elements were all converged.

At this point we will employ the Tikhonov regularization technique \cite{kruppa_george,tikhonov}.
The recipe that is followed was used for first time in the context of CSM in \cite{kruppa_george} and consists of three steps.
Initially the function that needs to be backrotated is mapped onto the interval (-$\infty$,+$\infty$) through the transformation:
\begin{equation}
\label{e_to_x}
f_{\theta}(x) = u_{\theta}(r_0e^{-x}),
\end{equation}
where u$_{\theta}$ is defined in Eq.\eqref{cs_wf} and $r_0$ = 1fm. The parameter $r_0$ serves the role of making 
variable $x$, as we will see below, dimensionless.
Then  Eq.\eqref{e_to_x} is Fourier transformed to obtain the:
\begin{equation}
\label{fourr}
u_{\theta}(\xi) = \frac{1}{\sqrt{2\pi}} \int_{-\infty}^{\infty}e^{-ix\xi}f_{\theta}(x)dx.
\end{equation}
Finally, the regularized backrotated wavefunction is computed as:
\begin{eqnarray}
\label{tikhonov}
u^{reg}(x+iy) = \frac{1}{\sqrt{2\pi}} \int_{-\infty}^{\infty}e^{-i(x+iy)\xi}   \nonumber   \\ 
\times \frac{u_{\theta}(\xi)}{1+\kappa e^{-2y\xi}} d\xi,
\end{eqnarray}
where $x$ = $\ln(r/r_0)$, y = $\theta$ and as we mentioned already  $x$ and also $\xi$ are dimensionless quantities.
The parameter $\kappa$ in Eq.\eqref{tikhonov} is the Tikhonov  parameter which controls
the smoothing of the backrotated wavefunction or the amount of regularization of  the inverse problem of backrotation in CSM.

\begin{figure}[t] 
  \includegraphics[width=\columnwidth]{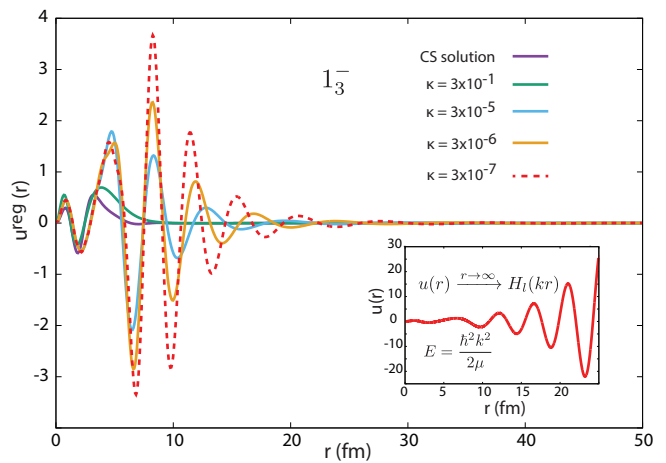} 
  \caption[T]{\label{fig3}
  (Color online)  Real parts of the reconstructed Gamow states for the 1$_{3}^{-}$ resonant state. Several states corresponding to a different regularization parameter are shown.
  In the inset it is shown the true $1^-_3$ resonant solution which is regular at the origin and has a pure outgoing behavior \cite{Vertse1982309}.}
\end{figure}

In Fig.\ref{fig3} we present the backrotated regularized $1_{3}^-$ wavefunction for several values of the parameter $\kappa$
and for comparison we also include the CSM $1_3^-$ wavefunction.
We notice that depending on the regularization parameter $\kappa$ we obtain different Gamow states. In practice a CSM solution once backrotated via the Tikhonov method will
generate an infinite number of Gamow states, since the parameter $\kappa$ can take any positive value. The wavefunctions are not observables so this feature is not
necessarily a problem, but what we will investigate later is what the impact of the parameter $\kappa$ and of the form of the Gamow function on an observable.

Another feature that we  observe is that the regularized wavefunctions do not exhibit the large amplitudes that we find in Fig.\ref{Fig2}.
We see that for different values of the parameter $\kappa$ what is changing is basically the asymptotic behavior of the backrotated wavefunction. 
A large parameter $\kappa$ is shown to over-regulate the backrotated wavefunction and the form looks similar to the CSM one e.g. for $\kappa$ = 3$\times$10$^{-1}$.
For distances r $<$ 3 fm (inside the potential well) the CSM wavefunction and the regularized backrotated wavefunctions look  almost identical.
For larger distances the regularized solutions are reminiscent of the behavior of an outgoing wave with an increasing amplitude for some distance outside
the potential ``lip" but with decaying amplitudes at  larger distances. By backrotating hence the CSM solutions, the Gamow character of the resonant state being an outgoing wave at large distances
is not fully retrieved. It is however interesting that for distances up to r $\sim$ 10 fm the behavior is more or else the  expected one, namely a bound-state like formation inside the potential
well and an outgoing wave just outside the well. It remains to be seen how observables will be affected once we will use the regularized backrotated states for the calculation of the expectation values.
In the inset of Fig.\ref{fig3} we are showing the behavior of the true Gamow solution with a complex energy $E = 2.017 -i\frac{0.972}{2}$MeV, which was calculated by integrating the Shr\"{odinger} equation subjected to pure outgoing boundary conditions \cite{Vertse1982309}. 

\subsubsection{Calculations of the root mean square radius}

Calculating an operator in a resonant state, as the one depicted in the inset of Fig.\ref{fig3}, will lead in divergent matrix elements and the $r^2$ operator is not an exception. 
Techniques such as the exterior complex scaling \cite{Gyarmati1971523} or Zel'dovich regularization \cite{Zel} are employed for the calculation 
of integrals. In a recent work for the description of rotational bands in $^8$Be, expectation values of transition operators in the continuum were calculated by adopting the Zel'dovich prescription \cite{garrido}, whereas in GSM for example \cite{review_GSM} diverging integrals are computed though the  exterior complex scaling.   

In the case of CSM, since the resonant state is always behaving as a bound state the divergence is not 
appearing (see Table \ref{tab:rad}). One may say that due to the discretization of the continuum in the HO basis, the integral for the $r^2$ calculation is regulated by the fixed b and fixed N of the HO truncation and when one is using the Gamow backrotated solutions the integral  is regulated via the Tikhonov method. 
Before continuing the discussion and applications of the backrotated wavefunctions, let us mention that the expectation value of an observable in a resonant state above threshold will always have
an imaginary part as we will notice for the rms radius. It was explained by Berggren that the physical interpretation of the imaginary part of the operator  is related to the uncertainty in the determination of its mean value \cite{Berggren19961}. 

We saw that the Hamiltonian operator in CSM is transformed under \eqref{cs_hami}. It is then expected that any quantum mechanical operator $\hat{O}$ will be transformed as:
\begin{equation}
\label{cs_op}
\hat{O}(\theta) = U(\theta)\hat{O}U(\theta)^{-1},
\end{equation}
 as for example in \textit{ab-initio} calculations of the dipole operator \cite{Horiuchi} or in benchmark CSM calculations \cite{kruppa_george,masui}.
In order to calculate its expectation value, one could use the transformed operator and calculate its expectation value between CSM solutions or calculate the expectation value of the bare operator but using the backrotated Gamow states.
Now that we have obtained the regularized backrotated solutions for several Tikhonov parameters we will calculate the root mean square (rms) expectation value of the radius square operator $r^2$. We will perform our calculations for the unrotated bare $r^2$ operator in the broad $1_{3}^-$ resonant state using the wavefunctions that
are shown in Fig.\ref{fig3} and the expectation value of the
complex scaled operator $r_{\theta}^2 \, = \,  U_{\theta} r^2 U_{\theta}^{-1}   \, = \, e^{2i\theta}r^2$  will be used as a benchmark. 

For the calculation we will use a large basis spanned by N=30 HO states so as to assure that $\langle r^2_{\theta} \rangle$
is fully converged as a function of basis states and as a function of $\theta$.   
For the rms radii calculations of  the $1_3^-$ resonant state we used a value of $\theta$ = 0.4 rad. The state is revealed at an angle $\theta$ $\sim$ 0.23 rad in the complex energy plane
and for values of $\theta$ = 0.35 and larger and for the size of the basis N=30 the radii have already converged up to the fourth significant digit as we present in Table \ref{tab:rad}.
\begin{table}[h!]
\caption{Dependence of the rms of the expectation value of the $r^2$ operator in the $1_3^-$ resonant state on the rotation angle $\theta$ for a basis of N=30 HO states.
Both the real and imaginary parts are shown and the radius is expresed as $\langle r^2\rangle^{1/2} = (\Re(\langle r^2\rangle^{1/2}),\Im(\langle r^2\rangle^{1/2}))$.  }
\begin{ruledtabular}
\begin{tabular}{cc}
\label{tab:rad}
$\theta$ (rad) &   $ \langle   r^2  \rangle^{\frac{1}{2}} $ (fm) \\
\hline
0.2 & (3.693,   1.763)  \\
0.3 & (3.227,  1.398) \\
0.4 & (3.220,  1.393) \\
0.5 & (3.220,  1.393) \\
0.6 & (3.220,  1.393) \\
0.7 & (3.220,  1.393) \\
\end{tabular}
\end{ruledtabular}
\end{table}

In Table \ref{tab:1} we gather the results for the rms expectation value of the $r^2$ operator for several renormalization parameters $\kappa$.
The tilde symbol  implies that conjugation does not affect the radial parts of the wavefunctions and stems from the fact that the solutions of the
complex scaled Hamiltonian do not satisfy the usual inner product but rather the generalized c-product \cite{moiseyev}. It is worth noting that the same holds for the backrotated
regularized solutions since they are also not part of the Hilbert space and they do satisfy the generalized c-product.
\begin{table}[h!]
\caption{Calculations of the rms of the expectation value of the $r^2$ operator in the regularized backrotated Gamow states.
Calculations correspond to N=30 and $\theta$ = 0.4. }
\begin{ruledtabular}
\begin{tabular}{cc}
\label{tab:1}
$\kappa$ &   $ \langle \widetilde{u^{reg}(r)}  | r^2 | u^{reg}(r) \rangle^{\frac{1}{2}} $ (fm) \\
\hline
3$\times$10$^{-1}$ & (4.36348,   0.80081)  \\
3$\times$10$^{-3}$ & (3.24315,  1.34967) \\
3$\times$10$^{-4}$ & (3.21818,  1.39037) \\
3$\times$10$^{-5}$ & (3.22016,  1.39346) \\
3$\times$10$^{-6}$ & (3.22010,  1.39341) \\
3$\times$10$^{-7}$ & (3.22009,  1.39345) \\
3$\times$10$^{-8}$ & (3.22005,  1.39359) \\
3$\times$10$^{-9}$ & (3.22018,  1.39323) \\
3$\times$10$^{-10}$ & (3.22003,  1.39456) \\
\hline
\hline \\
$\langle \widetilde{u_{\theta}(r)} | e^{2i\theta}r^2 | u_{\theta}(r) \rangle^{\frac{1}{2}}$  & (3.22008, 1.39351)
\end{tabular}
\end{ruledtabular}
\end{table}
For values of $\kappa$ smaller than 3$\times$10$^{-5}$ a plateau of rms radii starts to appear and the results coincide up to the fourth significant digit with the benchmark value of the CSM $r^2$ operator.
For values larger than 3$\times$10$^{-4}$ the result is deteriorating since then as was also shown in Fig.\ref{Fig2} there is an over-regulation of the backrotated Gamow state.
Overall, for a range of parameters $\kappa$ the rms radius computed from the backrotated wavefunctions is coinciding with the expectation value of the complex rotated operator and the fact that
the backrotated states go to zero for large distances, does not really affect the expectation value of the radius operator.  We can safely claim then that the Tikhonov backrotation produces a Gamow function which does not require 
extra care for treating the asymptotic distance divergence of a typical Gamow state, but at the same time reproduces the results that would be obtained using the exact outgoing solution of the Shr\"{o}dinger equation.

\subsubsection{Calculations of the dipole strength function.}

The calculation of the dipole strength function constitutes the calculation of the expectation value of the dipole operator between two different states. The so called dipole
transition is allowed between states with opposite parity and that differ by one unit of angular momentum. Testing the Tikhonov regularization for such an observable is a more 
stringent test since  the transition will involve an ensemble of states which will be backrotated and not a single state.  
The CSM allows to conveniently calculate the strength function via the complex scaled Green's function and by exploiting the completeness relation of the CSM resonant and and non-resonant scattering solutions.
The completeness relationship in CSM reassures that the many-body spectrum of the CSM Hamiltonian \eqref{cs_hami} which contains resonant and non-resonant  scattering states form a complete set \cite{Giraud2003115}
and allows to separate contributions of each element in computations of strength functions, phase-shifts and cross-sections \cite{Myo20141}. The completeness of the CSM spectrum is formally identical to the Berggren completeness \cite{Berggren1968265} relation which
was also utilized in the Gamow Shell Model (GSM) \cite{review_GSM}.

The dipole strength function $\mathcal{S}(E)$ in CSM then is given by the following formula:
\begin{equation}
\label{strength}
\mathcal{S}(E) = -\frac{1}{\pi} \sum_{\nu = 1}^{N}  Im \frac  {\langle \widetilde{  u_{\theta}^i(r)  } | \hat{O}(\theta) | u_{\theta}^{\nu}(r) \rangle \langle \widetilde{u_{\theta}^{\nu}(r)} | \hat{O}^{\dagger}(\theta) | u_{\theta}^{i}(r) \rangle }{E-E_{\nu}^{\theta}},
\end{equation}
where the indexes i and $\nu$ denote the initial and final states between which the transition occurs, $E$ is the excitation energy above threshold, $E_{\nu}$ are the complex
eigenstates of the CSM Hamiltonian which we do not distinguish at this point if they are resonant or non-resonant continuum states and $\nu$
then denotes the number of states we use for discretizing the continuum, hence it is equal to the number N of HO states we use in the basis. Usually, for calculations that
take place on the real energy axis, the number $\nu$ is also referred to as pseudostates for the fact that one describes the continuum in a bound state method spirit.
The operator $\hat{O}$ in our case is the dipole operator $\hat{O}$ = $\frac{4\pi}{3}rY_0^1$ which is  transformed as
$\hat{O}(\theta)$ = $\hat{O}$$e^{i\theta}$ and also $\hat{O}^{\dagger}(\theta)$ = $\hat{O}^{\dagger}$$e^{i\theta}$.
\begin{figure}[t] 
  \includegraphics[width=\columnwidth]{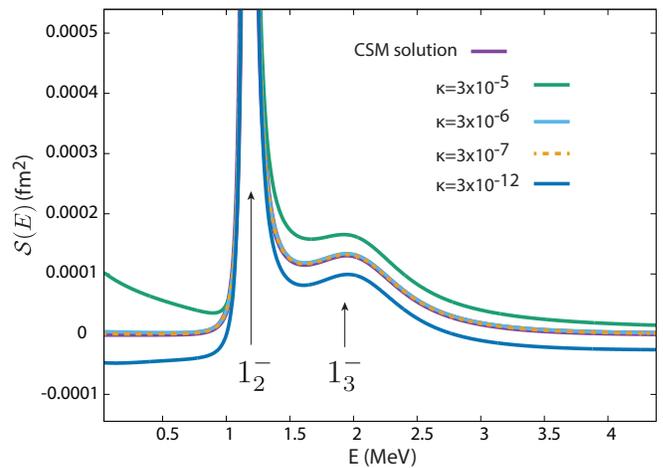} 
  \caption[T]{\label{fig4}
  (Color online)  Response function for the transition $0^+$ bound ground state $\to$ $1^-$ states in the continuum. The 1$^-$ CSM solutions were all backrotated using the Tikhonov method. For several Tikhonov parameters $\kappa$
  we compare the results against the CSM solution. The arrows denote the positions of the resonances (see also Fig.\ref{Fig1}).    }
\end{figure}
As we did for the radius operator, we backrotate in \eqref{strength} each one of the CSM solutions that are involved in the summation.
We then calculate the strength as:
\begin{equation}
\label{strength_br}
\mathcal{S}(E) = -\frac{1}{\pi} \sum_{\nu = 1}^{N}  Im \frac  {\langle \widetilde{  u_{reg}^i(r)  } | \hat{O} | u_{reg}^{\nu}(r) \rangle \langle \widetilde{u_{reg}^{\nu}(r)} | \hat{O}^{\dagger} | u_{reg}^{i}(r) \rangle }{E-E_{\nu}^{\theta}},
\end{equation}
using the bare dipole operator and not the $\hat{O}(\theta)$ one.
We gather our results on Fig.\ref{fig4}.
As we see, for a range of Tikhonov parameters the strength function \eqref{strength_br} calculated with the regularized Gamow states is identical to the strength function using formula \eqref{strength} and a similar situation was also encountered in rms radius calculation.
As a matter of fact, in \cite{kruppa_george} a criterion that eliminates too low and too large values of $\kappa$ was established, confirming that there is a plateau of regularization parameters for which results are converged. This is important to know  
as we may not have an ``exact" reference curve to our disposal.
Furthermore, we may also conclude that the regularized backrotated functions  also form a complete set which includes resonant and  non-resonant backrotated  states.

The problem of this chapter consisted of  studying the resonant features of two particles interacting in a potential well, which is very similar with the way s.p.  basis is generated for certain problems. We
have seen basis generating potentials of Woods-Saxon type \cite{Vertse1982309,review_GSM} or once again Gaussian potentials (e.g. KKNN \cite{kknn})  that are basically effective forces imitating the interaction of a target with a single projectile.
The possibility of obtaining a basis after backrotating CSM solutions could be considered and it
worths to check if the set of backrotated states forms an orthogonal set. 

\begin{figure}[t] 
  \includegraphics[width=\columnwidth]{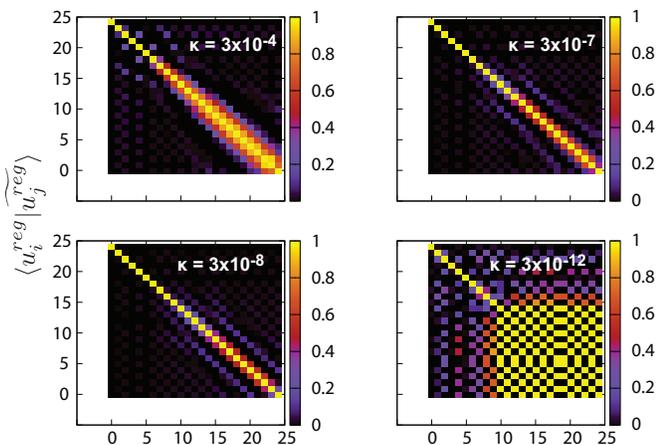} 
  \caption[T]{\label{fig5}
  (Color online)  Overlaps of the regularized backrotated $1^-$ states of the Gaussian schematic model for N=25 and $\theta$ = 0.3.    }
\end{figure}

In Fig.\ref{fig5} we plot the overlaps of the solutions for the  $1^-$ backrotated states.
We see that for $\kappa$ = 3$\times$10$^{-7}$ and $\kappa$ = 3$\times$10$^{-8}$, that  also belong in the plateau of parameters that provide excellent  agreement with the comparisons against the CSM solutions and the CSM operators, most of the overlaps are zero but the matrix is not strictly unity. The similarity with the unity matrix is
worse for other Tikhonov parameters.  In general the regularized backrotated states are not orthogonal and if one could find a way to utilize them in a basis expansion method that would
lead to a generalized diagonalization problem, with the calculation of the norm kernel being a necessity.

\subsubsection{Phase shifts and widths.}

We make a small parenthesis from studying the Tikhonov regularization in more realistic cases and we will compute the
scattering phase shifts produced by the schematic model. Within the CSM the resonance parameters are determined as the eigenstates of the CSM Hamiltonian matrix
which are stationary with respect to variations of the rotation angle.  An eigenstate then with complex energy $E = E_r - i\frac{\Gamma}{2}$ will  fully characterize the resonance.
Of course, resonance parameters can be also conveniently  calculated using techniques on the real energy axis. One of the most common ways to calculate resonance parameters on the real-axis
is by using the inflection criterion for the scattering phase shifts \cite{thompson}. The position of the resonance $E_r$ then is computed from the derivative of the phase-shift at the point that this derivative
is maximum and the widths are obtained using the formula: 
\begin{equation}
\label{inflection}
\Gamma = \frac{2}{d\delta(E)/dE} \big|_{E=E_r}.
\end{equation}
In CSM one can also calculate phase-shifts and then applying the inflection criterion formula we could see under which conditions are agreeing with each other.
\begin{figure}[t] 
  \includegraphics[width=\columnwidth]{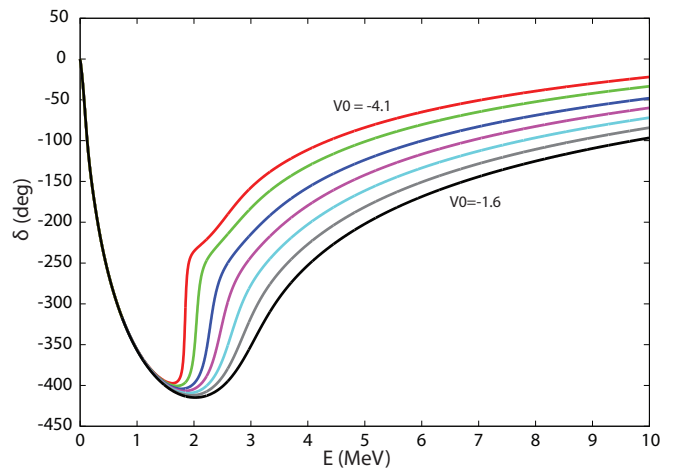} 
  \caption[T]{\label{fig6}
  (Color online)  Elastic scattering phase shifts for the schematic Gaussian potential as a function of the potential depth. The inflection criterion is applied to the phase shifts in order to deduce resonance parameters.
     See also Table {\ref{tab:2}}. }
\end{figure}

We calculate the phase shifts through the complex scaled continuum level density (CLD). The complex scaled CLD is defined as \cite{cld_csm}:
\begin{equation}
\label{cld}
\Delta^{\theta}(E) = -\frac{1}{\pi}Im \int d{\bf r} \langle {\bf r}| \frac{1}{E-H_{\theta}} - \frac{1}{E-H_{\theta}^{0}} | {\bf r^{\prime}} \rangle.
\end{equation} 
and
\begin{equation}
\delta(E) = \int_{0}^{E}\Delta^{\theta}(E)dE.
\end{equation}
In \eqref{cld} $H_{\theta}$ is the CSM interacting Hamiltonian whereas $H_{\theta}^{0}$ is the CSM kinetic energy. We should note that in \eqref{cld} all the eigenvalues of the complex
rotated interacting and asymptotic Hamiltonian are needed, nevertheless investigations on truncations of the number eigenvalues and the impact they have on the phase-shifts are underway.  
By calculating the CLD we could already determine resonance parameters because of its relation to the S-matrix and resonances appear as pronounced peaks on the CLD spectrum.
 The authors in \cite{kruppa_1998,kruppa_arai1,kruppa_arai2} showed that resonance parameters can be extracted by
calculating solely the CLD in an $L^2$ basis and after fitting the resonance region with Lorentzian or Breit-Wigner distributions.
Similar techniques were also used in the framework of Hartree-Fock-Bogoliubov (HFB) calculations \cite{hfb_qp} in order to extract resonance parameters from the HFB quasiparticle continuum space
of weakly bound nuclei. The only practical difference between calculating the CLD in a real energy formalism or within CSM is that in the CSM case one achieves a natural smoothing of the level density without
resorting to other smoothing techniques.
\begin{table}[h!]
\caption{ Resonances and widths for the potential $V(r) = -V_0e^{-0.16r^2} + 4.0e^{-0.04r^2}$ obtained from the diagonalization of $H_{\theta}$ directly and also from the scattering phase shifts using the R-matrix inflection criterion \eqref{inflection}. }
\begin{ruledtabular}
\begin{tabular}{ccc}
\label{tab:2}
$-V_0$ &  ($E_r, \Gamma$) diagonalization & ($E_r, \Gamma$) inflection \\
\hline
4.1 & (1.843, 5.343$\times$10$^{-2}$)  & (1.843, 5.422$\times$10$^{-2}$) \\
3.7 & (2.042, 0.118) & (2.046, 0.117) \\
3.2 & (2.279, 0.251) & (2.282, 0.240) \\
2.8 & (2.463, 0.402) & (2.471, 0.370) \\
2.4 & (2.644, 0.596) & (2.660, 0.522) \\
2.0 & (2.825, 0.834) & (2.857, 0.696) \\
1.6 & (3.009, 1.211) & (3.064, 0.886) \\
\end{tabular}
\end{ruledtabular}
\end{table}

The scattering phase-shifts in CSM can be evaluated by integrating \eqref{cld} over the range of energies. 
It has been shown that  coupled-channels can be treated in this way \cite{Suzuki_cc_cld}.   The relation that connects the CLD with the phase-shift is also encountered in
the work by \cite{ext_lev_theor,PhysRevA.22.101}
and it was generalized for the scattering of three-body systems (clusters) \cite{PhysRevC.40.1136}.
In Fig.\ref{fig6} we show the phase-shifts
calculated  for a two-body system with the particles interaction via the Gaussian potential \eqref{toy}. We changed the depth of the attractive form factor from -8.0 MeV to -4.1 MeV to obtain a single
resonance in the $\ell$=1 channel. In Table \ref{tab:2} we gather the resonance parameters for different potential depths as they are obtained from
the diagonalization of the CSM Hamiltonian matrix and also from the inflection criterion \eqref{inflection} for the phase-shift. 
We see that for resonances with a width as large as 600 keV the inflection formula and the result coming from the diagonalization of the complex matrix
are in good agreement. For broader resonances the inflection criterion is probably not so safe to use for extracting the width \cite{thompson}.
It is interesting however that the position of the resonance $E_r$ is in good agreement with the diagonalization result even in the case of broad resonances, so a conclusion that
can be drawn is that the inflection phase shift R-matrix criterion would work well for the description of resonances as broad as approximately 600 keV for the width,
but the position $E_r$ could be evaluated with good precision even when the resonance is much broader.

\subsection{Non-local potential used in CSM calculations and phase shifts}
When the potential has an analytical form in coordinate or momentum space, applications of the CSM transformation are trivial since the
CSM transformation can be directly applied to the coordinates of the potential.
Nowadays, most of the realistic microscopic potentials are given as matrix elements expressed in a HO basis in configuration space.
It  has been shown \cite{george1} that in this case the CSM can be also applied. We are repeating here the methodology that we followed.
Having an expression of HO basis matrix elements  which we will call $A_{nn^{\prime}}^{C;b}$ \cite{vary_heff_ls} 
 the following expansion is satisfied for the NN potential:
\begin{equation}
\label{sep_exp}
\hat{V_b} = \sum_{nn^{\prime};C} A_{nn^{\prime}}^{C;b} |n\rangle \langle n^{\prime} |
\end{equation}
where the quantum number $n$ denotes the 
nodes of the HO basis and $C$ denotes the specific channel which carries  the rest of the quantum numbers  and $b$ (or $\hbar \omega$) is the length parameter of the underlying HO basis.
From \eqref{sep_exp} one is able to express the potential in coordinate space as:
\begin{equation}
\label{sep_exp_coor}
V(r,r^{\prime}) = \langle r  | \hat{V_b} |  r^{\prime} \rangle = \sum_{nn^{\prime};C}  A_{nn^{\prime}}^{C;b} \phi_{n}^{C;b}(  r )\phi_{n^{\prime}}^{C;b}(  r^{\prime})
\end{equation}
where the $\phi$ functions stand for the analytical radial 3D HO wavefunctions.
It was numerically shown in \cite{george1} that treating the potential in this way also holds for a general class of potentials of non separable nature, such as chiral potentials.
Having the potential in this form we apply the CSM transformation and we do that in particular by shifting the CSM transformation from the potential to the HO basis, namely we calculate expressions of the form:
\begin{equation}
\label{cs_shift}
\begin{split}
V(re^{i\theta},r^{\prime}e^{i\theta}) &= e^{-i6\theta}\int_{0}^{\infty} \phi_n(re^{-i\theta}) \\
& V(r,r^{\prime})\phi_{n^{\prime}}(re^{-i\theta})r^2r^{\prime 2}drdr^{\prime}
\end{split}
\end{equation}
 which together with the complex scaled kinetic energy will lead to a complex symmetric eigenvalue problem and, in general, positions and widths of states above thresholds  and scattering phase shifts can be obtained. In \eqref{cs_shift},
 due to the analytical form of the HO basis one can either scale the coordinate $r$ or scale the HO length parameter $b$ $\to$ $be^{i\theta}$.
Notice that eventually the inverse CSM transformation on the HO basis will cause large oscillatory behavior for large $\theta$. Nevertheless, this oscillatory behavior does not cause any problem for rotation angles as large as $\theta$ $\sim$ 0.45 radians. 

\begin{figure}[t] 
  \includegraphics[width=\columnwidth]{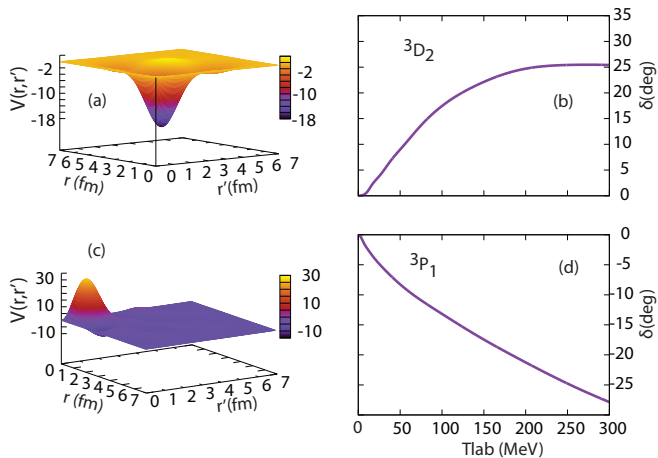} 
  \caption[T]{\label{fig7}
  (Color online)  Panels (a) and (c) : Non-local potentials produced using \eqref{sep_exp_coor} for the JISP16 interaction and then utilized within CSM to calculate scattering phase shifts as depicted in Panels (b) and (d).}
\end{figure}

\begin{figure}[h!] 
  \includegraphics[width=\columnwidth]{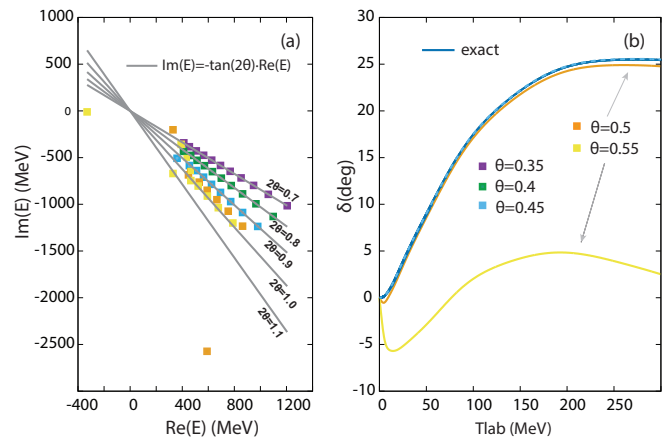} 
  \caption[T]{\label{fig8}
  (Color online)  Panel (a) : Part of the total $^3$D$_2$ eigenspectrum; it shows ten eigenvalues as a function of $\theta$ and as compared to the  $2\theta$ ray which the ABC theorem predicts. Panel (b) : The corresponding phase shifts which were computed by using the solutions in (a) are shown.}
\end{figure}

In all the following we employed the JISP16 NN realistic potential \cite{jisp16} at an $\hbar \omega$ = 30 MeV (or $b$ = 1.6627 fm). In Fig.\ref{fig7} we depict how the non local coordinate representations of the realistic JISP16 potential look like for the $^3$D$_2$ and $^3$P$_1$ channels. 
Also shown are phase shifts for the
corresponding channels, which are being calculated by utilizing the CSM solutions and the CLD formula \eqref{cld}.   
It is also of educational and illustrative purpose to notice the  attractive/repulsive nature of the interaction in these two different channels and
also the positive/negative phase shifts that are produced. From the numerical point of view, the phase shifts for $\theta$=0.4 radians and a number of HO radial nodes of N=20 ( see also Fig.\ref{fig8} have converged. We also performed calculations at different $\hbar \omega$ and checked convergence patterns with respect to N. 

The way one is able to obtain scattering phase shifts within the CSM while using a pure HO basis is an attractive characteristic. Methods that have employed an L$^2$ basis and
determined resonant and scattering characteristics of systems by varying the basis parameters are known in bibliography as L$^2$ stabilization methods \cite{stabil}. In this sense the HO basis CSM  can be seen as a complex analogue of an L$^2$ stabilization technique. Stabilization methods for the description of scattering exist in the field of Lattice QCD (LQCD) where the interacting particles are now positioned on a lattice instead of inside a HO. By varying the volume of the lattice, scattering phase shifts are then determined from the energy eigenstates  of the system \cite{silas_beane,dean_nature}. The mathematical connection between the volume dependence and the scattering observable is provided by the L\"{u}scher formula \cite{luscher} in LQCD  or the Busch formula in HO based calculations \cite{Busch,stetcu,Luu_2010}. The CSM also provides with a  way to describe scattering within an L$^2$ integrable basis, but at a lower computational cost \footnote{For the calculation of the phase-shifts in Fig.\ref{fig7} a rather modest HO basis was used, whereas when using the
Busch formula \cite{Luu_2010} a basis of N$\sim$1800 HO states was necessary. } and at the same time with the flexibility to be applied to the many-body scattering problem.  

It is useful at this point to make an investigation on the precision of the results  when varying the rotation angle $\theta$, discuss some current limitations and propose solutions for future applications.
First of all we would like to mention that the CSM with Gaussian potentials cannot be applied for rotation angles larger than $\sim$ 0.78 radians ($\pi/4$) since the potential starts to appear singularities for larger values,
so it would be useful to check the range of applicability when the interaction is realistic. We also refer the reader to an earlier work \cite{rittby1,*rittby2} where cases for rotation angles larger than the critical value of $\theta$ = $\pi/4$
where investigated.

It has been observed that when it comes to the phase-shifts (see also \cite{george1}), there is indeed a rapid convergence of the results to the exact phase shift \footnote{CSM should be seen as an
approximation of solving the Shr\"{o}dinger equation since it utilizes a finite set of basis states and also as a mathematical trick that turns the scattering problem into a bound state problem.  
As exact phase shift we consider a phase shift that stems from integrating the Shr\"{o}dinger equation with the correct asymptotic condition. } and the results are stable. We have noticed that the  stability is
 numerically related to the distribution
of the non-resonant scattering states along the $2\theta$ ray in the complex energy plane. If the distribution of  the non-resonant solutions is to a good approximation close to the
$2\theta$ ray, as it is also predicted by the ABC theorem, and the states do not depart much, then the phase shift calculated with \eqref{cld} is equivalent to the exact one. On the contrary if the non-resonant continua are scattered, then the quality
of the phase shift deteriorates. We show this behavior in Fig.\ref{fig8} where part of the solutions for the $^3$D$_2$ channel are shown and the corresponding phase shifts are calculated from this
spectrum.
Indeed, for rotation angles up to $\theta$=0.45 radians the calculated CSM phase shift is coinciding with the exact phase shift obtained by directly integrating the Shr\"{o}dinger equation.
Up to this point the complex eigenstates of the Hamiltonian (non-resonant continua) all fall almost exactly on a $2\theta$ path. For $\theta$ = 0.5 radians (orange points) the distribution of eigenstates
is departing from the 2$\theta$ path and so the phase shift becomes less accurate, departing from the exact one. Of course for $\theta$ = 0.55 radians (yellow points) the eigenstates are even more scattered and the 
phase shift is much different from the exact one. Even though for practical applications a rotation angle of $\theta$ $\sim$ 0.4 radians is
sufficient for revealing quite broad resonant states and also for convergence of observables, the stability issue for large rotation angles remains.
This is something that is well known within the CSM calculations that also have employed a HO basis and in our case is also related to the fact that in order to create 
complex scaled matrix elements of the NN interaction we shift the transformation to the HO basis, which as it was already mentioned causes a  large oscillatory behavior of the basis.  
It will be the topic of another publication to explore  more flexible basis but still in the framework of employing realistic NN potentials and also try to tackle more precisely the numerical integration
of functions that have large oscillatory behavior. It
is  also interesting to explore 
the possibility of generating pseudo-eigenstates that lie on a $2\theta$ trajectory and calculate phase shifts in this way. For phase shifts in particular it
is sufficient to only know the complex eigenstates for each channel without the need to calculate eigenvectors. Knowing already  the trajectory of these
solutions on the complex energy plane it would be worth trying to generate artificial non resonant eigenstates, as if they were produced by the eigensolver, and use the CLD formula to compute the phase shift.

\subsection{Backrotation application on deuteron for the dipole $^3$S$_1$-$^3$D$_1$ $\to$ $^3$P$_1$ transition  with a realistic force.}

We move forward to test the Tikhonov backrotation on the calculation the dipole transition from the deuteron g.s. to the $^3$P$_1$ continuum channel (see also test studies in a different context using the LIT \cite{leidemann}). 
At this point our goal is not to
benchmark the CSM method for calculating transitions with another method or compare with experiment, but we aim on testing the use of Tikhonov backrotated states for its calculation. 
Hence, for simplicity we have limited our selves to study the partial transition
only to the $^3$P$_1$ channel. Of course the Tikhonov method is a regularization technique that regularizes the backrotation transformation and there should
be no dependence of the regularization on whether the CSM solution comes from a toy model or a realistic Hamiltonian.
\begin{figure}[h!] 
  \includegraphics[width=\columnwidth]{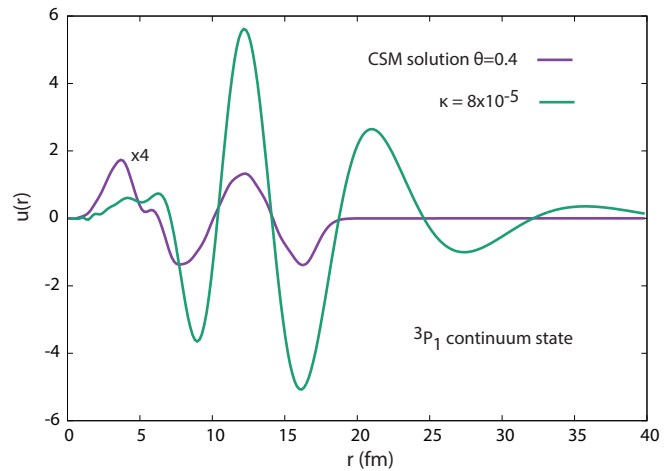} 
  \caption[T]{\label{fig9}
  (Color online)  Same as Fig.\ref{fig3} but for the $^3$P$_1$ non-resonant continuum wavefunction calculated with the JISP16 interaction. 
  What is shown is the real part of the radial behavior of the fourth normalized non-resonant continuum  $^3$P$_1$ solution that lies
  on the complex energy plane along a 2$\theta$ ray and $\theta$=0.4 radians. Also shown is the reconstructed  backrotated Gamow solution. For better visualization the CSM solution was multiplied by a  factor of 4.}
\end{figure}
We diagonalized the complex symmetric Hamiltonian matrix using a HO basis spanned by N=30 states and 
the CSM rotation angle was $\theta$ = 0.4 radians.   In Fig.\ref{fig9} we present the CSM solution for the  $^3$P$_1$ continuum channel, which in this case was the fourth continuum eigenstate.
What is also shown is the reconstruction of the Gamow backrotated wavefunction using the Tikhonov technique for a regularization parameter $\kappa$ = 8$\times$10$^{-5}$.
Then in Fig.\ref{fig10} we show the transition from the $^3$S$_1$-$^3$D$_1$ g.s. to the $^3$P$_1$ continua,  namely in Eq.\ref{strength_br} the
initial state is the deuteron g.s. and there is a sum over the HO basis complex scaled $^3$P$_1$ discretized continua which we have backrotated. As in the case of the Gaussian
potential, the response calculated using the CSM solutions and the complex scaled dipole operator served as a benchmark (see Eq.\ref{strength}).
We indeed observe that one can safely use the backrotated continua to calculate the response function with a more complicated and realistic interaction and there also exists a
plateau of Tikhonov parameters for which the result coincides with the benchmark CSM one. This is an indication that backrotated non-resonant continua
form a complete set in this case and actually the situation is the same with the one discussed around Fig.\ref{fig4}.
 
\section{Conclusions and perspectives}

In this work we have explored more aspects of the Tikhonov backrotation process for the calculation of observables in CSM.
In CSM even though one can conveniently determine resonant parameter of states above thresholds,
the resonant and non-resonant states are all expressed as linear combinations of L$^2$ integrable functions. Hence they exhibit an asymptotic behavior which is permitted within the CSM, but
is not the characteristic outgoing  asymptotic behavior of a resonant state. This is not affecting calculation of a large variety of expectation values of operators that can be easily complex rotated.
However it would always be useful within CSM to retrieve the correct asymptotic behavior of the Gamow wavefunction in calculations of excited states that are above particle thresholds. 
The Tikhonov method has proven to be suitable for this purpose.

We found out that the regularized backrotated Gamow wavefunction does not diverge at large distances which results in no special 
treatment for calculation of radial integrals. Even though the Gamow character was not fully retrieved this did not affect expectation values of observables in backrotated resonant states. 
This was shown by calculating expectation values of operators such as radii and response functions using the unrotated operator and the reconstructed Gamow functions.
The method was tested on a system of two particles interacting via a Gaussian potential and also in the deuteron for studying the
transition to the $^3$P$_1$ scattering state. We also investigated the orthogonality properties of the Gamow backrotated states for several regularization parameters and found out that
they are not orthogonal.
\begin{figure}[h!] 
  \includegraphics[width=\columnwidth]{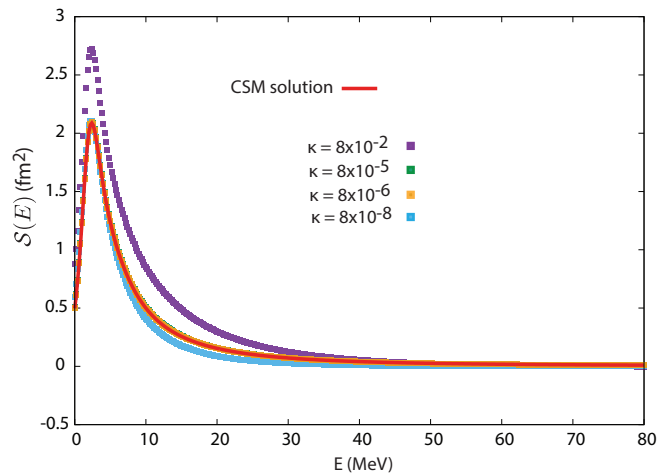} 
  \caption[T]{\label{fig10}
  (Color online)  Same as Fig.\ref{fig4} but for the transition from the deuteron g.s. to the $^3$P$_1$ continua and also using the realistic JISP16 interaction.}
\end{figure}

At the same time we explored some other features of the CSM, such as the ability to have access to both resonant parameters and also
scattering quantities such as  phase shifts. Within the schematic Gaussian model we were able to compare resonant parameters stemming
from the CSM diagonalization and the scattering phase shift using the so-called inflection criterion. The inflection formula provided good results for resonances
as broad as 600 keV whereas for broader resonances a complex energy method such as the CSM appears to be more accurate.

We studied the behavior of the CSM scattering eigenvalues for large rotation angles. For $\theta$ as large as 0.4 radians the scattering
states are distributed along the, expected by the ABC theorem, 2$\theta$ line. We found out that this is the necessary condition to have stable and converged scattering 
phase shifts. Increasing the rotation angle to very large values causes the departure of the solutions from the 2$\theta$ line and at the same time the phase shift becomes
unstable. 

This problem would be treated by employing  different L$^2$ basis sets  for the discretization of the continuum which may increase precision of calculations and even use different complex energy eigenvalue
solvers. In particular it is of interest to us to invest time on implementing L$^2$ basis for the discretization of the continuum in CSM that are defined on a grid such as Lagrange basis \cite{Baye20151} or wavelet basis \cite{accurate_pots}, since it was recently shown that they can increase the 
precision, in particular the distribution of continuum states along the 2$\theta$ cuts of the complex energy plane. 
We would also like to use more advanced quadratures to integrate matrix elements between backrotated HO states or alternatively to  regularize via the Tikhonov method the HO wavefunctions entering \eqref{cs_shift} and then use these states to express the complex scaled interaction matrix elements and
check if results would stabilize for angles larger than $\theta$ $\sim$ 0.5 radians. 

\begin{acknowledgments}
 This work was prepared by LLNL under Contract No.  DE-AC52-07NA27344.    
 This  material  is  based  upon  work  supported by the U.S. Department of Energy, Office of Science, 
Office of Nuclear Physics, under Work Proposal No.
SCW0498-59806 and Award Number DE-FG02-96ER40985. 
This work was also supported in part by the US DOE under grants No. DESC0008485 (SciDAC/NUCLEI) and DE-FG02- 87ER40371.
We would like to thank A. T. Kruppa, W. Nazarewicz and J. P. Vary for discussions on the topic and comments on the paper and also  J. P. Vary for sharing his  realistic NN interaction code. We would like to thank A. T. Kruppa who pointed  out the possibility of using the Tikhonov method for the backrotation task,
 T. Vertse for his assistance with the code GAMOW and R. Lazauskas for discussions and for pointing out the work of Rittby $\textit{et al}$.
Part of this work  originated and completed during the workshops ``International Collaborations in Nuclear Theory:
Theory for open-shell nuclei near the limits of stability" and  ``Computational Advances in Nuclear and Hadron Physics"   that took place at Michigan State University and the Yukawa Institute for Theoretical Physics respectively. We would like to thank the organizers for the hospitality.
\end{acknowledgments}

\bibliographystyle{apsrev4-1}
\bibliography{CS_tikh}    

\end{document}